\documentclass[%
12pt,
superscriptaddress,
reprint,
 amsmath,amssymb,
 aps,
prb,
]{revtex4-2}
\usepackage{amsmath, amsthm, amssymb, amsfonts}
\usepackage{float}
\usepackage{tikzsymbols}
\usepackage[export]{adjustbox}
\usepackage{xcolor}

\usepackage[caption=false]{subfig}

\usepackage{ragged2e}

\DeclareCaptionOption{justified}[]{\caption@setjustification{justified}}

\usepackage{enumitem}
\usepackage{graphicx}
\usepackage{comment}
\graphicspath{ {Figs/} }

\usepackage{physics}
\usepackage[colorlinks,allcolors=cyan]{hyperref}
\usepackage[capitalise]{cleveref}
\usepackage[T1]{fontenc}
\usepackage[english]{babel}

\begin{document}
\title{Collective interlayer pairing and pair superfluidity in vertically stacked layers of dipolar excitons}
\author{Michal Zimmerman }
\affiliation{The Racah Institute of Physics, The Hebrew University of Jerusalem, Jerusalem 9190401, Israel}
\author{Ronen Rapaport }
\affiliation{The Racah Institute of Physics, The Hebrew University of Jerusalem, Jerusalem 9190401, Israel}
\author{Snir Gazit }
\affiliation{The Racah Institute of Physics, The Hebrew University of Jerusalem, Jerusalem 9190401, Israel}
\affiliation{The Fritz Haber Research Center for Molecular Dynamics,
The Hebrew University of Jerusalem, Jerusalem 9190401, Israel}

\begin{abstract}
Layered bosonic dipolar fluids have been suggested to host a condensate of interlayer molecular bound states. However, its experimental observation has remained elusive. Motivated by two recent experimental works [Hubert et al., Phys. Rev. X 9, 021026 (2019) and D. J. Choksy et al., Phys. Rev. B 103 045126 (2021)], we theoretically study, using numerically exact quantum Monte Carlo calculations, the experimental signatures of collective interlayer pairing in vertically stacked indirect exciton (IX) layers. We find that IX energy shifts associated with each layer evolve non trivially as a function of density imbalance following a nonmonotonic trend with a jump discontinuity at density balance, identified with the interlayer IX molecule gap. This behavior discriminates between the superfluidity of interlayer bound pairs and independent dipole condensation in distinct layers. Considering finite temperature and finite density imbalance conditions, we find a cascade of Berezinskii--Kosterlitz--Thouless (BKT) transitions, initially into a pair superfluid and only then, at lower temperatures, into complete superfluidity of both layers. Our results may provide a theoretical interpretation of existing experimental observations in GaAs double quantum well (DQW) bilayer structures. Furthermore, to optimize the visibility of pairing dynamics in future studies, we present an analysis suggesting realistic experimental settings in GaAs and transition metal dichalcogenide (TMD) bilayer DQW heterostructures where collective interlayer pairing and pair superfluidity can be clearly observed.
\end{abstract}

\maketitle

\begin{figure*}[t]
    \captionsetup[subfigure]{labelformat=empty}
	\subfloat[\label{fig:Fig_1a}]{}
    \subfloat[\label{fig:Fig_1b}]{}
    \subfloat[\label{fig:Fig_1c}]{}
    \subfloat[\label{fig:Fig_1d}]{}
    \includegraphics[width=\textwidth]{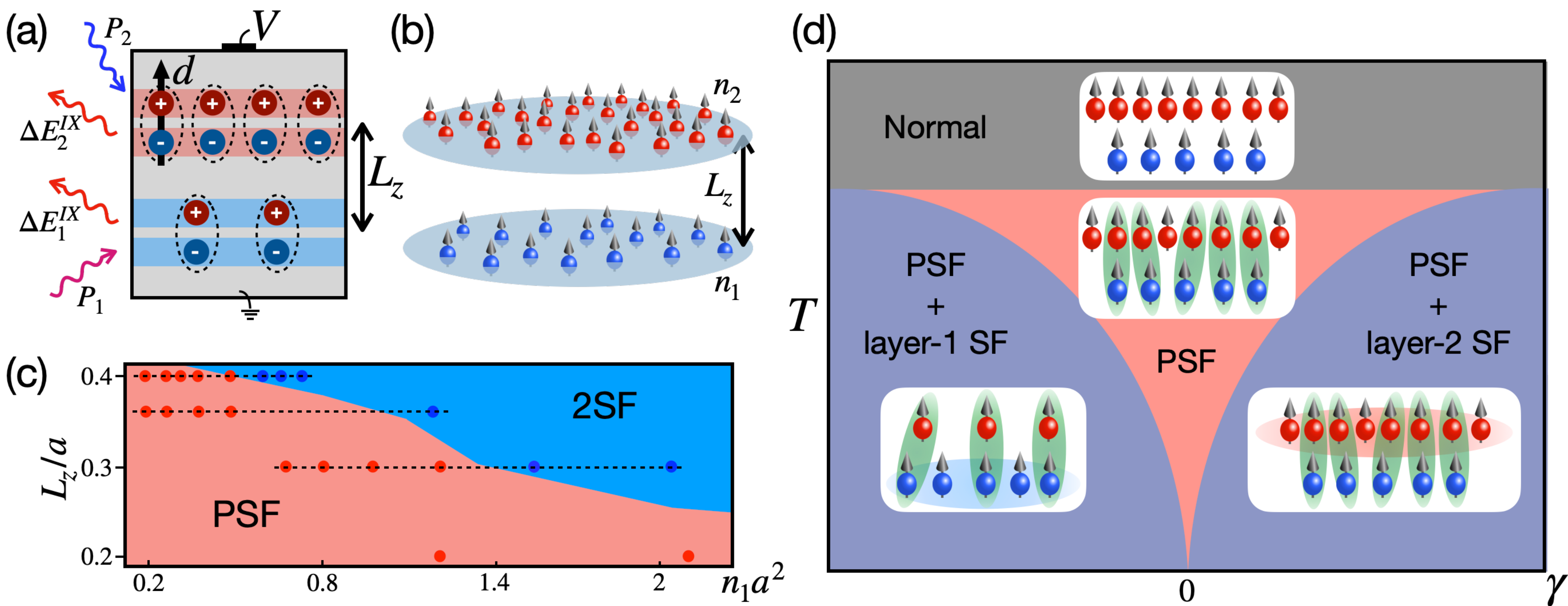}
    \caption{(a) Double quantum well bilayer heterostructure. The bottom and top layers, colored in blue and red, respectively, are vertically separated by a distance of $L_z$. Externally applied voltage $V$ creates spatial separation between electrons (blue circles) and holes (red circles) within each layer, resulting in IXs with fixed dipole size $d$. Pump powers $P_\alpha$ control the IX densities, while photons emitted during IX recombination are used to measure IX energy shifts $\Delta E_\alpha^{\text{IX}}$ within each layer. (b) Effective model of the experimental system (a), consisting of dipolar bosons in a bilayer geometry with possibly distinct densities $n_1,n_2$. The interlayer separation is denoted by $L_z$. (c) Equal density ($n_1=n_2$) ground state phase diagram as a function of $L_z/a$ and IX density $n_1a^2$. Red (blue) dots correspond to numerical data points in the PSF (2SF) phase, and black dashed lines are guides to the eye for the parameter cuts considered in this work. (d) Phase diagram at finite temperature $T$ and finite imbalance $\gamma$, emanating from the zero temperature and vanishing density imbalance ($\gamma=0$) PSF phase.}
    \label{fig:Fig1}
\end{figure*} 

\section{Introduction}
\label{sec:introduction}

Pair superfluids (PSF) are molecular Bose condensates made out of tightly bound pairs of the underlying elementary bosonic degrees of freedom \cite{Nozi_1982}. Convenient settings for stabilizing PSF phases are strongly correlated Bose mixtures \cite{Altman_2003,Kuklov_2004}, where interspecies attractive interactions promote the formation of a bound pair, and intraspecies repulsion protects from a real-space Bose collapse. Remarkably, PSFs are predicted to support highly nontrivial physical phenomena, such as half-vortex configurations \cite{Lee_1985,Carpenter_1989} and enhanced inter-species superfluid drag \cite{Andreev_1975,Sellin_2018}, which further motivates their experimental realization.

An intriguing theoretical proposal \cite{Trefzger_2009} for realizing a PSF is a bilayer geometry comprising bosonic particles that are confined to propagate in the plane and carry a fixed dipole moment (either electric or magnetic), aligned perpendicular to the layers. The short range attractive component of the interlayer dipolar coupling supports an interlayer bound state \cite{Klawunn_2010,Cohen_2016}, which in turn may condense and form a PSF. This scenario was confirmed in numerical simulations of lattice models \cite{Safavi_Naini_2013} and in the continuum \cite{Macia_2014,Cinti_2017}.

While the idea of realizing a PSF phase in dipolar bilayers has a long theoretical history, it has remained elusive experimentally. Motivated by experimental breakthroughs in stabilizing electric \cite{Christoph_2014,Simon_2014,Wang_2016} and magnetic \cite{Benjamin_2011,Ferlaino_2012} dipolar condensates in cold atoms, most theoretical proposals have focused on such systems. However, experimental challenges such as suppressing two-body losses \cite{Jun_2021} and reaching interlayer binding energies which are sufficiently large compared to the accessible temperature scales \cite{Kruckenhauser_2020}, still remain. 

In that regard, excitons in two-dimensional landscapes have in recent years emerged as the solid state analog of cold atoms in the quest of realizing novel quantum many-body phases. In particular, experimental advances in transition metal dichalcogenide (TMD) materials paved the route for observing collective quantum states, including Mott insulators \cite{Xu_2020,Regan_2020,Wang_2020}, Wigner crystals \cite{Regan_2020}, stripe phases \cite{Jin_2021} and an excitonic insulator \cite{Gu_2022}. 

Excitonic systems have several advantages, including the ability to precisely confine excitons into deep subwavelength potential planes and study mutual interactions with other types of quantum matter, such as low-dimensional electron gases, via proximity coupling \cite{Demler_2021,Shimazaki_2021}. Moreover, exciton densities are readily controlled via external excitation sources, and interexciton correlations can be directly inferred from their spectral shifts \cite{Laikhtman_2009}, giving a direct probe of collective phases and phase transitions. 

More concretely, electrically induced excitons (IXs) in two-dimensional DQW structures, either GaAs- or TMD-based, have recently attracted a great deal of interest as a promising platform for observing collective phenomena of interacting quantum dipolar liquids \cite{Lozovik1996,Butov_2002,Combescot_2012,Shilo_2013,Fogler_2014,Dubin_2017,Misra_2018,Wang_2019,Mazuz_2019,Slobodkin_2020,Holleitner_2020,Lagoin_2021,Misra_2021,Gu_2022}. Considering multilayered DQW heterostructures is a particularly interesting research avenue, as it enables revealing the anisotropic and attractive component of dipolar interactions, beyond the purely repulsive dipolar interaction within a monolayer. This line of research is further motivated by recent observations of exotic quantum states in cold atoms carrying magnetic dipoles, such as dipolar instabilities and the formation of the elusive supersolid phase \cite{Tanzi_2019,Tilman_2019,Chomaz_2019}, all of which arise from the anisotropic nature of dipolar interactions in three dimensions. 

The simplest DQW multilayered layout is a bilayer geometry at a fixed and well-defined lateral separation $L_z$, which was recently realized in GaAs heterostructures \cite{Cohen_2016,Rapaport_2019,Choksy_2021} (\cref{fig:Fig_1a}). In the experimental setting, IX densities at each layer ($\alpha=1,2$) are controlled independently in a continuous fashion via the laser pump power, $P_\alpha$. This crucial experimental tuning knob allows one to continuously probe finite imbalances between layer densities. The experimentally measured photoluminescence, resulting from IX recombination processes, measures the energy released during the annihilation of a single IX. Interactions between IXs are then encoded in spectral shifts $\Delta E_\alpha^{\text{IX}}$, and at zero temperature are identified with the IX chemical potential $\Delta E_\alpha^{\text{IX}}=\mu_\alpha=E(N_\alpha+1)-E(N_\alpha)$ \cite{Laikhtman_2009}, with $N_\alpha$ denoting the number of IXs at the $\alpha$'s layer.

Both of the aforementioned experiments have observed a nontrivial behavior of IX energy shifts as a function of pump power and associated IX densities. Specifically, Ref.~\cite{Rapaport_2019} observed a nonmonotonic evolution of IX energies in one of the layers as a function of the density in the complementary one. Ref.~\cite{Choksy_2021} has reported a monotonic red (blue) shift of the probed layer as a function of laser pumping power applied to the complementary (same) layer. The appearance of red shifted IX energies in both experiments provides experimental evidence for the effective coupling and the energy gain associated with interlayer dipolar attraction, which inspires exploring experimental signatures of the resulting quantum many-body phenomena.

In this work, we show that the unique advantages of stacked layers of dipolar excitons enable realizing and directly probing the emergence of collective paired dynamics. To that end, we chart the low-temperature phase diagram in the presence of finite density imbalance, using quantum Monte Carlo (QMC) simulations of an effective bosonic model of vertically stacked IXs. We determine the IX energy shifts, $\Delta E_\alpha^{\text{IX}}$, associated with each layer and uncover a highly nontrivial evolution as a function of microscopic parameters and density imbalance. Importantly, we argue that measurements of $\Delta E_\alpha^{\text{IX}}$ allow one to experimentally discriminate between PSF and independent IX condensates in each layer. Most notably, in the PSF phase, we predict a nonmonotonic behavior of the spectral shift as a function of density imbalance and a sharp jump discontinuity in $\Delta E_\alpha^{\text{IX}}$ at density balance, identified with the binding energy of a  molecule of two IXs, one from each layer. Our findings shed light on existing experimental results and serve as a guide for ongoing and future experimental studies of bilayer IXs in GaAs and TMD heterostructures, for which we propose explicit experimental parameters and protocols.


\section*{Effective Model and Phase Diagram}
 
\label{sec:eff_model}
  
We model the bilayer IX system via point-like dipolar bosons confined in two vertically separated planes at a fixed separation $L_z$ (\cref{fig:Fig_1b}). The bosonic approximation is appropriate in the dilute limit, where the spatial extent of the electron hole bound state is significantly smaller than the inter-IX separation. The resulting model Hamiltonian comprises a sum over the kinetic energy term in each layer (interlayer hopping is disallowed) and intralayer and interlayer dipolar interactions. Explicitly,

\begin{equation}\label{eqn:H_dipole}
H=-\sum_{\alpha,i_{\alpha}} \frac{\hbar^2\nabla^2_{i_\alpha}} {2m} + \sum_{\alpha,i_{\alpha}< i'_{\alpha}} U_{dd}(\vb{r}_{i_{\alpha}},\vb{r}_{i'_{\alpha}})+\sum_{i_{1},i_{2}} U_{dd}(\vb{r}_{i_{1}},\vb{r}_{i_{2}}).
\end{equation}

Here $\alpha=1,2$ marks the layer index, $i_\alpha$ labels different IXs at each respective layer, and $\vb{r}_{i_\alpha}$ denotes their locations. We take the same effective mass, $m$, for both layers and set IX densities to $n_\alpha=N_\alpha/A$, where $A$ is the area. Two body interactions, in the last two terms, are given by the dipolar potential,
\begin{equation}\label{eqn:U_dipole}
U_{dd}\left(\vb{r},\vb{r}'\right) = D^2\,\frac{\abs{\vb{r_{\bot}}-\vb{r'_{\bot}}}^{2}-2\abs{z-z'}^{2}}{\left(\abs{\vb{r_{\bot}}-\vb{r'_{\bot}}}^{2} + \abs{z-z'}^{2}\right)^{\frac{5}{2}}}.
\end{equation}

Here $\vb{r_{\bot}}$ and $z$ denote in-plane and vertical positions of the IXs, respectively. The dipole strength equals $D^2=e^2 d^2 / 4\pi \varepsilon_r \varepsilon_0$, with $e$ being the electron charge, $d$ being the dipole size, and $\varepsilon_0$ ($\varepsilon_r$) being the vacuum (relative) permittivity. We measure all length scales with the effective dipolar length $a=\frac{mD^2}{\hbar^2}$ and energy scales via $E_0=\frac{D^2}{a^3}$ and define the density imbalance $\gamma=(n_2-n_1)/n_1$ induced by varying $n_2$, while keeping $n_1$ fixed.
We note that $a=149,46$ nm and $E_0=0.016,0.073$ meV for dipole size $d=22,4$ nm, corresponding to GaAs and TMD based structures, respectively. See further details below. 
From a symmetry perspective, the Hamiltonian has a global $U(1)\times U(1)$ symmetry, corresponding to independent boson particle number conservation at each layer.

We begin our analysis by reviewing the zero temperature phase diagram in the density balanced case, $\gamma=0$ \cite{Macia_2014,Cinti_2017} (\cref{fig:Fig_1c}). We focus on liquid phases pertinent to the experimentally dilute IX limit. For large layer separations, $L_z \gg a$, the two layers decouple, giving rise to independent superfluids (2SF) at each layer \cite{Zoller_2007,Lozovik_2007}. In the opposite limit $L_z \ll a$, the strong attractive component of dipolar interactions promotes the formation of interlayer dipolar bound pairs \cite{Cohen_2016,Klawunn_2010}, whose condensate forms a PSF. 
As a function of layer separation, the PSF phase remains stable at higher densities for smaller layer separations due to the stronger interlayer binding.
We note that for certain fixed $L_z$ values, the phase transition between the PSF and 2SF phases can also be tuned via the IX density $n_1(=n_2)$. This observation is crucial from the experimental perspective since IX densities, as opposed to the fixed sample depended lateral separation $L_z$, are continuously varied parameters, readily controlled by the excitation source pump power.

Importantly, the PSF phase only partially breaks the global $U(1)\times U(1)$ down to $U(1)$ \cite{Altman_2003,Kuklov_2004} by assigning a non vanishing expectation value for the interlayer pair annihilation operator, $\mathcal{O}_{\text{PSF}}=b_1 b_2$, $\expval{\mathcal{O}_{\text{PSF}}}\ne 0$. The remaining conserved $U(1)$ symmetry corresponds to the super-counter-fluid (SCF) channel, whose associated order parameter $\mathcal{O}_{\text{SCF}}=b_1 b^\dagger_2$ generates density imbalance and hence leads to breaking of interlayer pairs. The energy penalty associated with pair breaking, in the PSF phase, endows a finite gap to the SCF channel \cite{Macia_2014}. By contrast, in the 2SF phase, both layers condense independently, rendering both the PSF and SCF channels gapless, i.e. $\expval{\mathcal{O}_{\text{PSF}}}\ne 0$ and $\expval{\mathcal{O}_{\text{SCF}}}\ne 0$.  

\begin{figure}[t]
    \captionsetup[subfigure]{labelformat=empty}
	\subfloat[\label{fig:Fig_2a}]{}
    \subfloat[\label{fig:Fig_2b}]{}
    \includegraphics[width=0.48\textwidth,left]{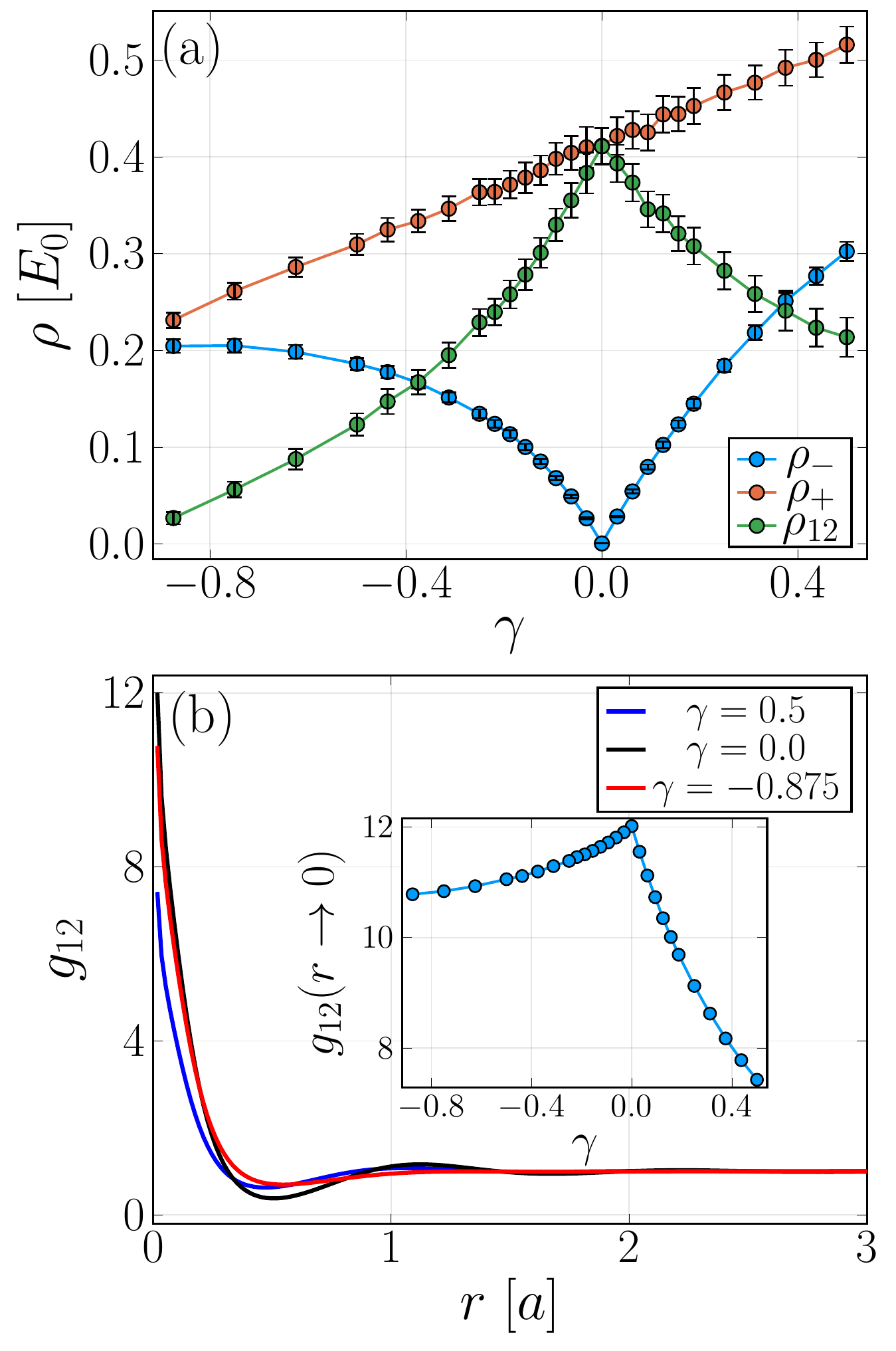}
    \caption{
   The influence of density imbalance on the PSF phase at zero temperature, for $L_z=0.3[a]$ and $n_1=0.83[a^{-2}]$. (a) The superfluid stiffness in the SCF channel (blue) and PSF channel (red) and the superfluid drag (green) as a function of $\gamma$. (b) Interlayer density-density correlation function $g_{12}(r)$ for various density imbalances. (Inset) $g_{12}(r\to0)$ as a function of $\gamma$. The above results were obtained at temperature $T=0.08[E_0]$ and system size $N_1=32$.
    }
    \vspace{-1.5em}
    \label{fig:Fig2}
\end{figure} 

We now turn to the main focus of this work, which is to establish the role of density imbalance on the global phase diagram and determine its experimental signatures. At strictly zero temperature, starting from the PSF state, even the slightest density imbalance is expected to nucleate an excess of unpaired bosons, which will condense at layer $\alpha=1(2)$ for $\gamma<0$  $(\gamma>0)$. The resulting phase, labeled by PSF+layer-1(2) SF, has an identical symmetry breaking ($U(1)\times U(1)$) as the 2SF phase and hence can not be distinguished based on symmetry probes. Therefore, the SCF channel, which was originally gapped at density balance in the PSF phase, is now also condensed.

This singular behavior is depicted in the zero temperature cut of the phase diagram shown in \cref{fig:Fig_1d}, where the PSF phase is restricted to a single point at $\gamma=0$. Consequently, experimentally targeting the PSF phase and directly probing its properties is challenging as it would require a high degree of fine tuning in order to precisely equate the layer densities. Nevertheless, below, we will argue that the ability to optically induce and control finite IX density imbalances is, in fact, a feature enabling probing pairing dynamics.

Away from the zero temperature limit, following the Mermin--Wagner theorem, thermal fluctuations destroy the long-range off-diagonal order. Nevertheless, at sufficiently low temperatures, quasi-long-range order with power law correlations survives. The short-range correlated normal fluid state then appears at higher temperatures, above the BKT transition temperature, $T_{\text{BKT}}$. Concerning our problem, a key insight is that the BKT transition temperature is proportional to the density of the condensed constituents. Consequently, for small density imbalance, the minority of excess particles participating in the SCF condensate suppresses the associated BKT temperature, giving rise to a cascade of BKT transitions \cite{Kuklov_2004} obeying $T_{\text{BKT}}^{\text{PSF}}>T_{\text{BKT}}^{\text{SCF}}$. The resulting quasi-long-range ordered PSF phase develops a fan structure at finite temperatures, as depicted in \cref{fig:Fig_1d}. At temperatures lower than $T_{\text{BKT}}^{\text{SCF}}$, the PSF and the single superfluid in layer $\alpha=1(2)$ for $\gamma<0(\gamma>0)$ coexist, forming a PSF+layer-1(2) SF phase.

Previous studies of imbalanced IX bilayers were treated analytically in the polaron limit, $N_1=1$ \cite{Rapaport_2020,Chao_2021}, or numerically via variational techniques in \cite{Chao_2021}, primarily in the layer-independent superfluid phase. These calculations are approximate in nature and have limited predictive power in the strongly correlated regime of the full two-dimensional problem at arbitrary IX densities and strong pairing. In the following, we turn to numerically establish the global phase diagram beyond the above limiting cases and determine the physical signatures of various phases and phase transitions in the context of IX bilayer experiments carried out at finite density imbalance, focusing on physical properties of the interlayer pair condensate.  

\begin{figure}[t]
    \captionsetup[subfigure]{labelformat=empty}
	\subfloat[\label{fig:Fig_3a}]{}
    \subfloat[\label{fig:Fig_3b}]{}
    \includegraphics[width=0.48\textwidth,left]{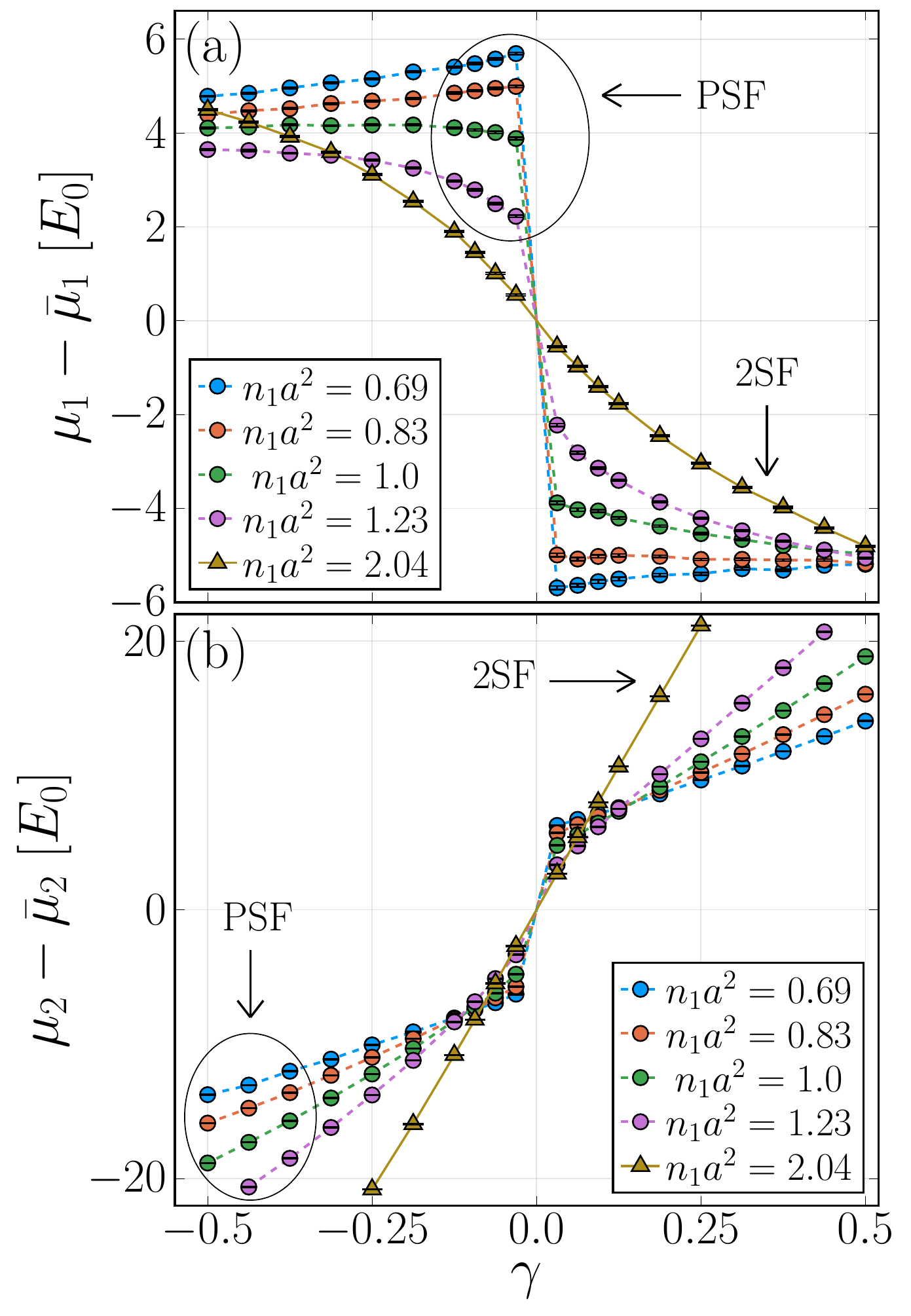}
    \caption{
    Evolution of the IX chemical potential, $\mu_\alpha$, corresponding to (a) layer $\alpha=1$ and (b) layer $\alpha=2$, as a function of density imbalance at $L_z=0.3[a]$, $T=0.3[E_0]$ and $N_1=32$. Dashed and solid lines correspond to curves that are either PSF or 2SF at equal densities, respectively. Marking circles and pointing arrows are used to further emphasize the two phases. Vertical shifts $\bar{\mu}_\alpha$ are defined such that the plotted curves cross the origin at $\gamma=0$.
    }
    \vspace{-1.5em}
    \label{fig:Fig3}
\end{figure} 

\section*{Numerical Methods and Observables}
\label{sec:methods}

To numerically study the Hamiltonian in \cref{eqn:H_dipole}, we employ path integral Monte Carlo calculations using efficient worm algorithm updates \cite{Prokof'ev_2006}. We consider the finite temperature formulation and address the ground state physics by tracking the convergence of our finite temperature data to the zero temperature limit; see, e.g., \cref{app:qmc}. 

To detect the presence of PSF or SCF or both, we compute the associated superfluid stiffness. The PSF (SCF) is manifest by long range correlations of the sum (difference) of phases, $\theta_\pm=\theta_1\pm\theta_2$, where we identify $b_\alpha\sim e^{i\theta_\alpha}$ for weak density fluctuations \cite{Kuklov_2004}. For multispecies bosonic systems, the superfluid stiffness is a tensor which measures the response to minimally coupled global flux insertions, $\Phi_\alpha$, for each species, $\rho_{\alpha,\alpha'}=\frac{\partial^2 F}{\partial \Phi_\alpha \partial \Phi_{\alpha'}}$, where $F$ denotes the free energy. The superfluid stiffness in the PSF (SCF) channel is then obtained by taking the flux configuration $\Phi_1=\Phi_2=\Phi/2$ ($\Phi_1=-\Phi_2=\Phi/2$). Within the world-line representation, both quantities can be measured via the standard winding number variance $\rho_{\pm}=\frac{\expval{(\vb{W}_1\pm \vb{W}_2)^2}}{4\beta \mathcal{D}}$ \cite{Pollock_1987,Svistunov_2015}. Here $\vb{W}_\alpha=\{W^x_\alpha,W^y_\alpha\}$ denotes the winding numbers along the $x/y$ directions in the layer $\alpha$, the inverse temperature is $\beta=1/T$, and $\mathcal{D}=2$ is the spatial dimension. A related observable is the superfluid drag $\rho_{12}=\frac{\expval{\vb{W}_1\cdot \vb{W}_2}}{\beta \mathcal{D}}$, which quantifies the response of a given layer to a superfluid flow in the complementary one, also known as the Andreev-Bashkin effect \cite{Andreev_1975,Sellin_2018}.

To quantify spatial correlations, we compute the density-density correlation function $g_{\alpha,\alpha'}(\vb{r})= \frac{A}{N_\alpha N_{\alpha'}}\expval{\sum_{i_\alpha,j_{\alpha'} \neq i_\alpha}\delta(\vb{r}-\vb{r}_{i_\alpha} + \vb{r}_{j_{\alpha'}})}$, for both the intralayer, $\alpha=\alpha'$, and interlayer, $\alpha\ne\alpha'$, cases. Last, to numerically estimate the experimentally measured IX energy shifts, $\Delta E_\alpha^{\text{IX}}$, we monitor the finite temperature chemical potential of each species $\mu_\alpha=F(N_\alpha+1,T)-F(N_\alpha,T)$. To that end, we employ the technique proposed in Ref.~\cite{Herdman_2014}

\begin{figure}[t]
    \captionsetup[subfigure]{labelformat=empty}
	\subfloat[\label{fig:Fig_4a}]{}
    \subfloat[\label{fig:Fig_4b}]{}
    \includegraphics[width=0.48\textwidth,left]{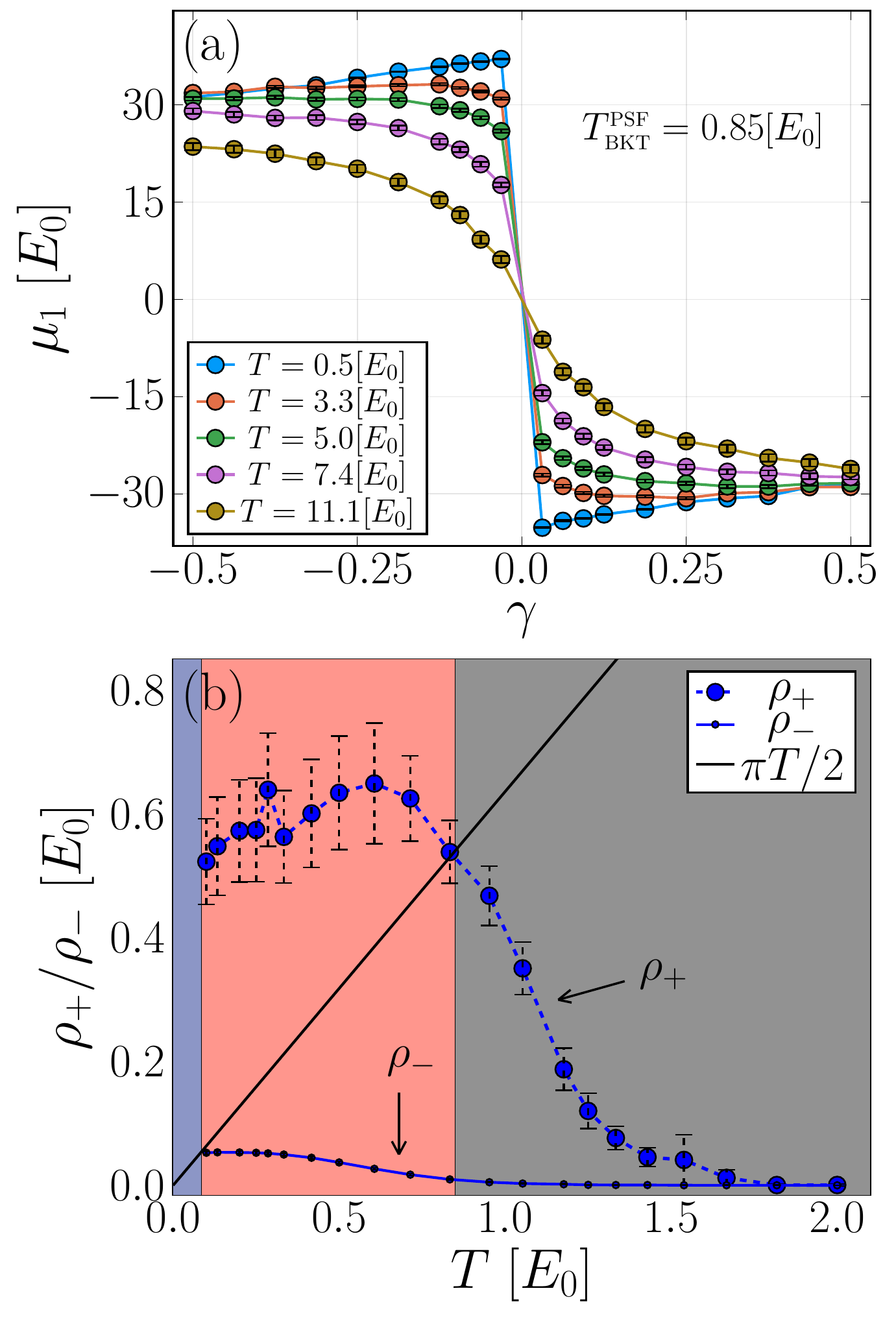}
    \caption{(a) $\mu_1$ as a function of $\gamma$ at finite temperatures for $n_1=1.23[a^{-2}]$ and $L_z=0.2[a]$. As a reference, the BKT transition temperature of the PSF phase for the corresponding set of parameters equals $T_{\text{BKT}}^{\text{PSF}}=0.85[E_0]$. (b) Superfluid stiffness channels as a function of temperature for the same ($L_z$, $n_1$) as in (a) and fixed $\gamma=0.125$. The arrows point to the curves of $\rho_+$ and $\rho_-$, and the black line cuts each stiffness at the corresponding critical temperature for the BKT transition, given by the Nelson relation. We take  $N_1=32,48$ in (a) and (b), respectively. 
    \vspace{-1.5em}}
    \label{fig:Fig4}
\end{figure} 

\section*{Numerical Results } 
\label{sec:results}

As concrete microscopic values, pertinent to the experimental parameters to be discussed below, we consider several values of interlayer separation $L_z/a=0.4,0.36,0.3,0.2$, for which we study a range of IX densities $n_1[a^{-2}]$ along the parameter cut connecting the PSF and 2SF phases in the density-balanced phase diagram, marked by dashed lines in \cref{fig:Fig_1c}. For each parameter pair ($L_z$, $n_1$), we consider a range of density imbalances, $\gamma$, by varying $n_2$ while retaining $n_1$ fixed.  We present results up to $N_1=48$ and down to $T=0.08[E_0]$ for ground state properties. See finite size and finite temperature analysis in \cref{app:qmc}.

We start our analysis by determining the fate of the paired condensate away from $\gamma=0$. We fix $L_z=0.3[a]$ and $n_1=0.83[a^{-2}]$, such that the ground state in the density-balanced case is a PSF phase. In \cref{fig:Fig_2a}, we examine the superfluid stiffness in the PSF and SCF channels, $\rho_{\pm}$, as a function of density imbalance. We find that $\rho_+$ remains finite irrespective of $\gamma$. On the other hand, $\rho_-$ vanishes at $\gamma=0$ and rises continuously away from that point. These results corroborate the above reasoning, indicating that an arbitrarily small excess of unpaired IXs condenses at sufficiently low temperatures, breaking the full $U(1)\times U(1)$ symmetry. 

To investigate interlayer correlations at finite density imbalance, we first compute the superfluid drag coefficient $\rho_{12}$ in \cref{fig:Fig_2a}. We observe that $\rho_{12}$ is maximal at the balanced point and decreases as the system is tuned towards greater imbalance. The presence of a sizable superfluid drag suggests a correlated motion of interlayer IX pairs, even away from the balanced density point. To further study interlayer pairing in this parameter regime, we examine the interlayer density-density correlation function, $g_{12}(r)$, depicted in \cref{fig:Fig_2b}. We indeed observe strong interlayer correlation, manifest in a peak structure of $g_{12}(r)$ at zero horizontal separation, even at a finite imbalance. The peak's height gradually decreases as the imbalance grows, as can be seen in the nonmonotonic behavior of $g_{12}(r\to0)$ as a function of $\gamma$, presented in the inset of \cref{fig:Fig_2b}. Further results on the evolution of the interlayer and intralayer density-density correlations are presented in \cref{app:qmc}.

We now turn to compute our main observable, which is the chemical potential $\mu_\alpha$, identified with the experimentally measured IX energy shifts $\Delta E_{\alpha}^{\text{IX}}$. In \cref{fig:Fig_3a}, we depict $\mu_1$ at $L_z=0.3[a]$ as a function of density imbalance for several densities $n_1$. To facilitate the comparison between distinct densities, we introduce a vertical shift, $\bar{\mu}_1$, such that all curves cross the origin at the density balanced point $\gamma=0$. We first focus on the dilute limit, corresponding to the PSF phase for $\gamma=0$, by setting $n_1=0.69 [a^{-2}]$. We observe that for negative imbalance, $\gamma<0$, $\mu_1$ increases as a function of $\gamma$. 

To understand the above behavior, we note that, for strong pairing, all IXs at layer $\alpha=2$ form tightly bound pairs with IXs at the complementary layer $\alpha=1$. Thus, for $\gamma<0$, any additional IX at layer $\alpha=1$ will lack a pairing partner at layer $\alpha=2$. Consequently, adding a single IX to layer $\alpha=1$ will incur an additional energy penalty, beyond the usual intralayer repulsion, arising from the repulsive long-distance tail of interlayer interactions. Increasing the density of IXs in layer $\alpha=2$ will further enhance this interlayer repulsion, as deduced from the functional form of the interlayer potential in the dilute limit. The above reasoning gives rise to the observed increase in $\mu_1$ as a function of $\gamma$.

On the other hand, for $\gamma>0$, at least one IX in layer $\alpha=2$ is unpaired, such that adding an IX at layer $\alpha=1$ will gain molecular binding energy. This pairing is evident in the jump discontinuity $\Delta \mu_1=|\mu_1(\gamma=0^+)-\mu_1(\gamma=0^-)|$ seen at $\gamma=0$. Physically, in the dilute density limit, the size of the jump equals the binding gap of a bound interlayer IX molecule \cite{Macia_2014,Cohen_2016}. Similar to the $\gamma<0$ case, further enlarging $\gamma$ results in an increase of $\mu_1$. To see why this is the case, we note that besides the bound pair, all other IXs in layer $\alpha=2$ and remaining bound pairs repel the added IX in layer $\alpha=1$, resulting in an energy penalty and an increase of the chemical potential $\mu_1$. Remarkably, from the above analysis, we conclude that $\mu_1$ curves exhibit a nonmonotonic evolution as a function of $\gamma$ in the dilute limit, which can potentially explain the experimental observations of Ref.~\cite{Rapaport_2019}. More broadly, the functional form of $\Delta E_1^{\text{IX}}$ serves as a clear experimental fingerprint of interlayer pairing. 

Tracking the evolution of $\mu_1$ for several $n_1$ values connecting the PSF and 2SF phases, in \cref{fig:Fig_3a} we observe that the jump discontinuity in $\mu_1$ progressively vanishes upon approach to the 2SF phase. The softening of this energy scale marks the position of the associated quantum critical point separating the two phases \cite{Macia_2014}. We analyze the associated universal properties in \cref{app:qmc}. Furthermore, $\mu_1$ curves transition between a nonmonotonic trend at low densities to a monotonic decrease at high densities. The latter behavior was observed experimentally in Ref.~\cite{Choksy_2021}. Crucially, this transition can be probed via a continuous tuning of layers densities, allowing for direct experimental observation of quantum critical dynamics. Quantitatively similar results were obtained for all other $L_z$ values.

Turning our attention to $\mu_2$ (\cref{fig:Fig_3b}), we observe a monotonic increase as a function $\gamma$, indicating the thermodynamic stability of this phase, as was observed in Ref.~\cite{Choksy_2021}. Beyond the general trend, similarly to $\mu_1$, we find a non-analytic jump seen precisely at $\gamma=0$. As before, the size of the jump $\Delta \mu_{2}=|\mu_{2}(\gamma=0^+)-\mu_{2}(\gamma=0^-)|$ corresponds to the IX molecule binding energy for dilute layer densities and vanishes as we transition to the 2SF phase by increasing the IX density.

Similarly to the calculations in Refs.~\cite{Rapaport_2020,Chao_2021}, obtained in the polaron limit ($\gamma \gg 1$), we find that, at least for small densities, $\mu_1$ monotonically increases with $\gamma$ (see dashed blue line in \cref{fig:Fig_3a}). At higher density values, the considered parameter regime is likely away from the polaron limit, making the comparison more subtle.

Next, we examine finite temperature effects on the measurement of IX energy shifts for $L_z=0.2[a]$ and $n_1=1.23[a^{-2}]$. In \cref{fig:Fig_4a}, we depict the chemical potential $\mu_1$ as a function of density imbalance for a set of increasing temperatures. The activated behavior controlled by the molecular IX pairing gap smears the jump discontinuity, which eventually gives rise to a smooth monotonic decrease in $\mu_1$ at sufficiently high temperatures. 

Last, we pin down the BKT temperatures associated with the PSF and SCF channels using the standard superfluid stiffness jump analysis. This is achieved by locating the intersection between the superfluid stiffness and the linear curve $\frac{2}{\pi} T $ \cite{Nelson_1977}. The results of this analysis are shown in \cref{fig:Fig_4b} for $L_z=0.2[a]$, $n_1=1.23[a^{-2}]$ and $\gamma=0.125$. Indeed, we observe a clear separation of transition temperatures, with $T_{\text{BKT}}^{\text{PSF}}>T_{\text{BKT}}^{\text{SCF}}$, which generates the predicted fan like structure of the PSF phase appearing in \cref{fig:Fig_1d}. A similar analysis was used to determine the PSF BKT temperature for all other microscopic parameters. See \cref{app:qmc} for a finite size scaling analysis of the BKT transition.

\begin{figure}[t]
    \captionsetup[subfigure]{labelformat=empty}
	\subfloat[\label{fig:Fig_5a}]{}
    \subfloat[\label{fig:Fig_5b}]{}
    \includegraphics[width=0.45\textwidth]{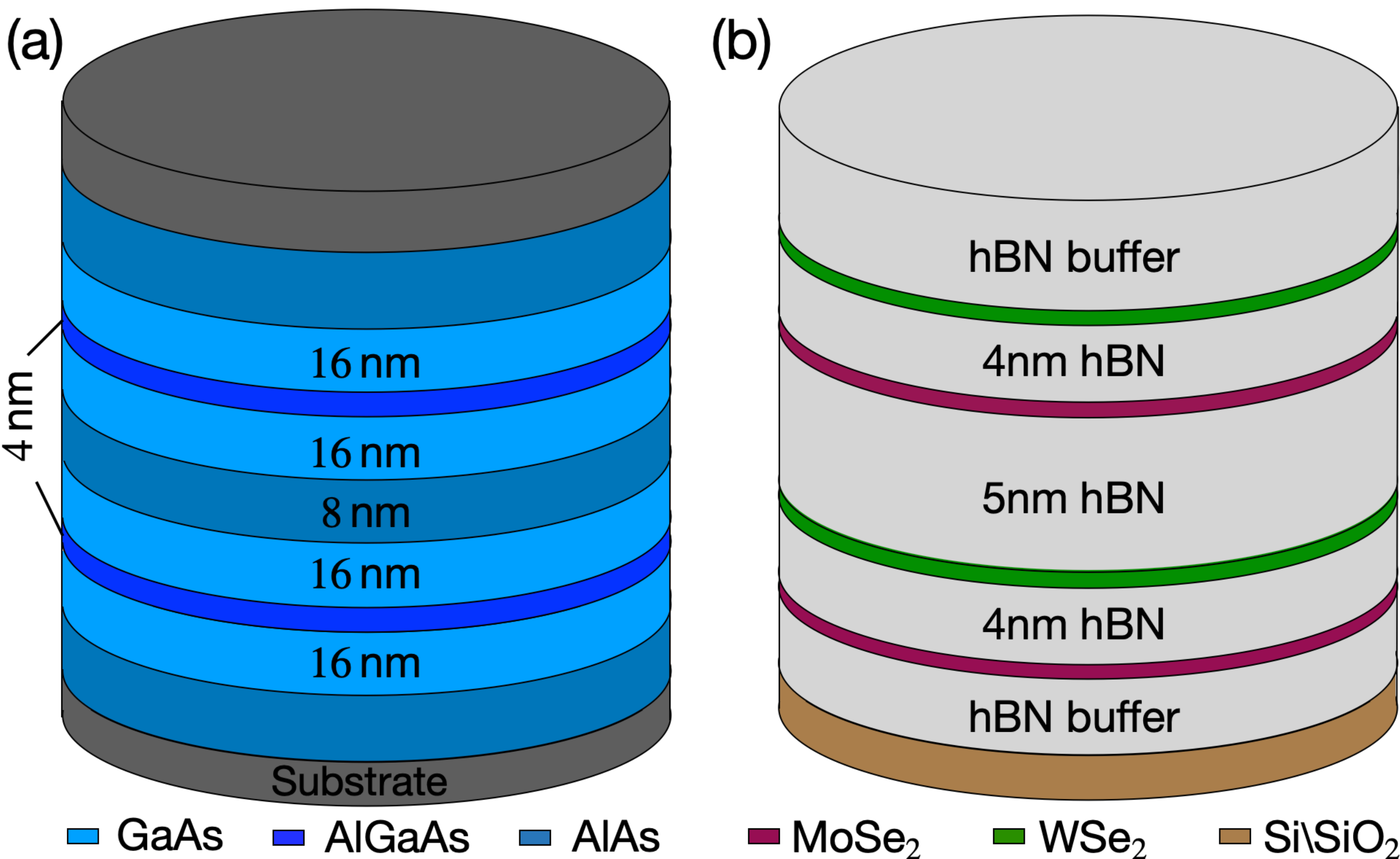}
    \vspace{0.5em}
    \caption{Experimental proposal for DQW bilayer structures. (a) GaAs: Each DQW layer $\alpha$ comprises two $16$-nm GaAs quantum wells with a $4$-nm Al$_{0.4}$Ga$_{0.6}$As barrier. An $8$-nm AlAs separates the different layers and prevents interlayer tunneling. The sample is grown on a doped substrate and has a top gate, which allows the application of an external voltage along the growth direction. (b) TMD: A single layer $\alpha$ is defined by MoSe$_2$ and WSe$_2$ monolayers, with a $4$-nm hBN buffer in between. An additional $5$-nm hBN barrier separates the two MoSe$_2$/hBN/WSe$_2$ structures. 
    \vspace{-1.5em}
    }
    \label{fig:Fig5}
\end{figure} 
\section*{Experimental Considerations } 
\label{sec:exp_considerations}


In this section, we quantitatively discuss the experimental visibility of our predictions in GaAs and TMD DQW bilayer structures. We focus on layer symmetric structures, see \cref{fig:Fig5}, and fix the ratio between the dipole size and layer separation to $d/L_z=0.5$ and $0.45$ for GaAs and TMD, respectively. Since the effective mass and relative permittivity are fixed material properties, the dipole size $d$ and IX densities are the only remaining experimental tuning knobs, which we optimize to enhance the experimental signatures of paired dynamics.

The most important energy scales for experimentally characterizing the PSF phase are the IX molecule binding gap $\Delta \mu_{1,2}$ and the PSF BKT temperature, $T_{\text{BKT}}^{\text{PSF}}$. The binding gap saturates to its largest value in the dilute IX limit and vanishes for higher IX densities upon approaching the 2SF phase. The BKT temperature, on the other hand, scales linearly with the density and hence is maximized in the opposite limit of high IX densities. To resolve this trade-off, we target the largest possible IX densities for which the IX molecule binding gap remains relatively close to its dilute limit maximal value.

With the above reasoning in mind, we first examine GaAs heterostructures, see \cref{fig:Fig_5a}, similar to the ones studied in \cite{Rapaport_2019,Choksy_2021}.
Explicitly, each DQW layer consists of a pair of $16$-nm GaAs quantum wells separated by a $4$-nm Al$_{0.4}$Ga$_{0.6}$As barrier. The vertically stacked DQWs are then split up by an $8$-nm AlAs barrier preventing inter-DQW tunneling. We take effective IX mass of $m=0.21m_e$, with $m_e$ being the free electron mass, and relative permittivity $\varepsilon=12.9$. To determine the dipole size, we numerically solved the single-particle Schr\"{o}dinger equation for the electron and hole (\cref{app:IX}) for the above setting. Assuming a uniform cross-section and bias electric field of $F_e=-2.5$ V/$\mu$m yields an effective dipole size $d=22$ nm. The inter-layer separation, measured as the distance between the centers of the lower and upper DQWs, then equals $L_z=44$ nm, such that, using the definitions given above, we find the dimensionless ratio $L_z/a=0.3$, as studied numerically above. 

With the above experimental parameters, for IX densities in the range $n_1\approx3-9\times 10^{9}$ cm$^{-2}$, we estimate the jump discontinuity in the IX energy shift to be $\Delta E^{\text{IX}}_\alpha \approx0.02-0.19$ meV, where the magnitude of $\Delta E^{\text{IX}}_\alpha$ increases as the IX density decreases (\cref{app:qmc}). Targeting for the lowest IX density at which the onset of pair superfluidity is feasible at realistic experimental temperature scales, we suggest considering IX density $n_1\approx3.7\times 10^{9}$ cm$^{-2}$, for which the IX energy shift equals $\Delta E^{\text{IX}}_\alpha \approx0.16$ meV and interlayer pair condensation is expected to occur at temperatures lower than $T_{\text{BKT}}^{\text{PSF}}= 0.11$ K. 

\begin{figure}[t]
    \includegraphics[width=0.45\textwidth]{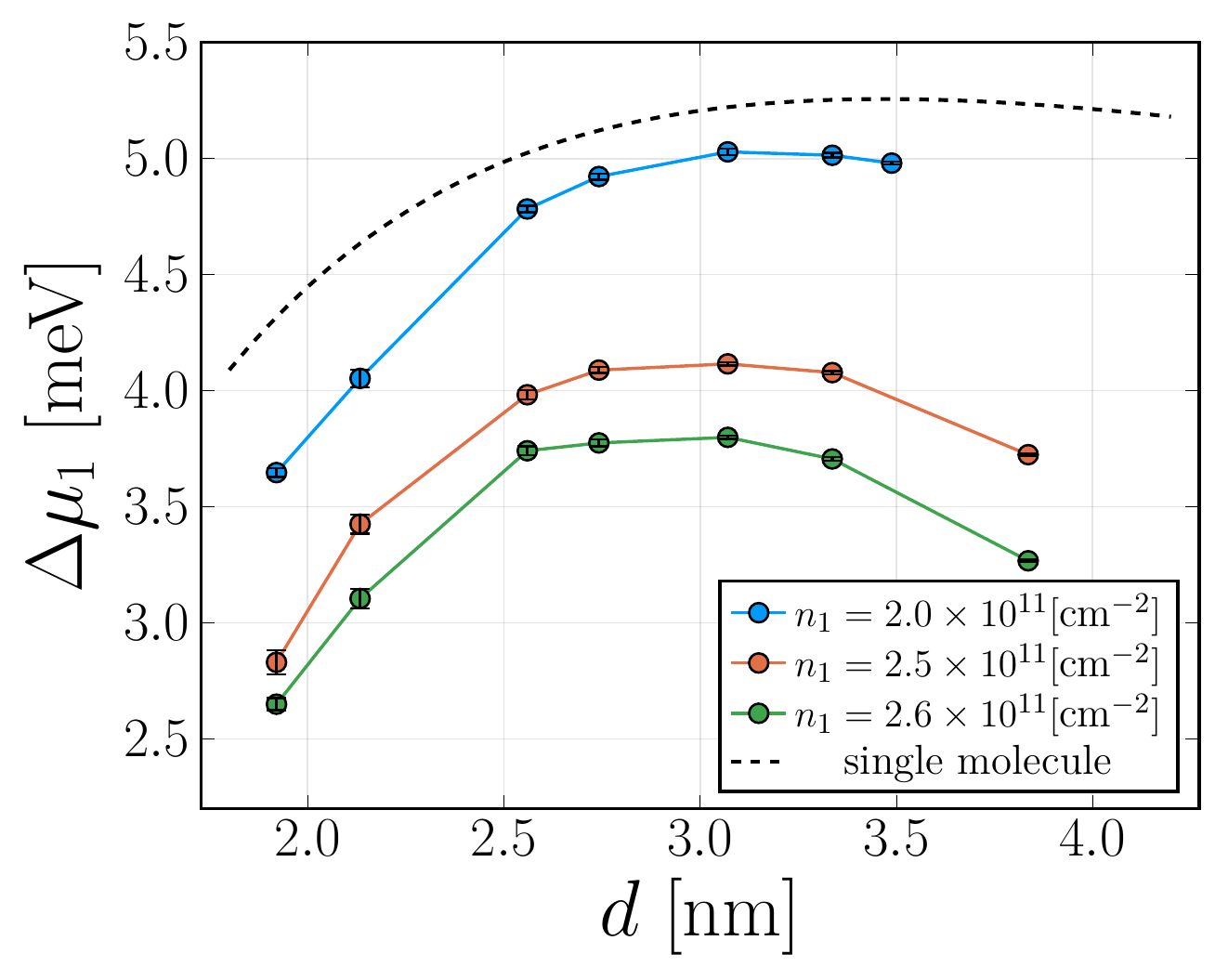}
    \vspace{1.5em}
    \caption{The pair binding gap $\Delta\mu_1$ as a function of dipolar size for several $n_1$ values and a fixed ratio $d/L_z=0.45$. We consider typical parameters of TMD materials: $\varepsilon=3.3$, $m=0.5m_e$. The data points were obtained for the ground state limit with $N_1=16$. The dashed line shows the binding energies of a single IX molecule, representative of the extreme dilute limit.
    \vspace{-1.5em}}
    \label{fig:Fig6}
\end{figure} 

TMD structures comprised of MoSe$_2$/We$_2$ monolayers with an hBN buffer \cite{Calman_2018,Lee_2020} offer a particularly promising setting for the experimental realization and detection of PSFs. Compared to GaAs, TMDs are characterized by large effective mass $m=0.5m_e$ and low permittivity $\varepsilon=3.3$, giving rise to significantly larger dipolar length, $a$, as defined above. Crucially, TMD structures can potentially enable experimental access to the strong binding regime.

As a concrete realization, we suggest a bilayer stacking of MoSe$_2$/hBN/WSe$_2$ DQW structures, separated by an hBN buffer to prevent IX tunneling between layers. Specifically, we propose an IX dipole size of $d=4$ nm, set by $12$ hBN layers \cite{Kumashiro_2000} sandwiched between MoSe$_2$ and We$_2$ monolayers. The two DQW layers are then further separated by $15$ hBN layers, resulting in $L_z=9$ nm (\cref{fig:Fig_5b}). In dimensionless parameters, we obtain $L_z/a=0.2$, which, for a wide range of densities, is significantly deeper in the PSF phase.

For the above setting, we estimate IX energy shifts $\Delta E^{\text{IX}}_\alpha\approx 3.5-5$ meV for IX densities in the range $n_1\approx 6-19\times 10^{10}$ cm$^{-2}$, with a decreasing trend in $\Delta E^{\text{IX}}$ with increasing density. More specifically, for the highest IX density $n_1\approx 9.7\times 10^{10}$ cm$^{-2}$ at which the jump in the IX energy shifts saturates (\cref{app:qmc}) on $\Delta E^{\text{IX}}_\alpha\approx 5$ meV, we expect a condensation of the PSF channel at $T_{\text{BKT}}^{\text{PSF}}\approx1.18$ K. Crucially, TMD structures present significantly larger binding energies and condensation temperatures, which greatly facilitate the experimental realization and detection of paired phases compared to GaAs systems. 

Although, in practice, there is limited flexibility in tuning the dipole size $d$, it is interesting to understand how to optimize $d$ in order to maximize $\Delta\mu_1$ for a given IX density. To that end, we examine the evolution of $\Delta\mu_1$ as a function of $d$ in a TMD-based system for several densities $n_1$ (\cref{fig:Fig6}). We fix the ratio $d/L_z=0.45$, such that, geometrically, increasing $d$ amounts to a linear scaling of the entire sample, including $L_z$. Interestingly, we find that $\Delta\mu_1$ exhibits a non-monotonic behavior as a function of $d$. 

The above result can be traced back to the simple interlayer two-body problem, which approximates the many-body behavior in the dilute limit \cite{Cohen_2016}. The corresponding binding energies, obtained from the solution of the single IX molecule Schr\"{o}dinger equation, are shown by a dashed line in \cref{fig:Fig6}. Constraining a fixed $d/L_z$ ratio, the only remaining length scale in the problem is the dipole size $d$. Using simple dimensional analysis, one can show that the interlayer attractive dipolar potential scales as $U_{dd}\sim 1/d$, whereas the kinetic energy term scales as $E_{K}\sim 1/d^2$. Hence, for sufficiently small dipole sizes, the kinetic energy dominates, leading to a decrease in the binding energy due to delocalization. In the opposite limit of large dipole sizes, the reduction in the potential energy again leads to diminished binding energies. The competition between these trends resolves in a nonmonotonic evolution of the binding energy as a function of dipole size.

As a final comment, we note that in the strong binding regime and for low densities, there is a clear separation between the binding energies $\Delta \mu_{1,2}$ and the pair BKT temperature $T_{\text{BKT}}^{\text{PSF}}$. Consequently, at intermediate temperatures, $T_{\text{BKT}}^{\text{PSF}}<T<\Delta \mu_{1,2}$, the IX energy shift $\Delta E^{\text{IX}}_1$ is still expected to exhibit a sharp jump at density balance, albeit slightly rounded due to finite temperature effects. Physically, the resulting phase is identified with a normal liquid of paired interlayer molecules. The paired liquid will eventually condense only at lower temperatures, below the BKT transition. From the above reasoning, identification of the PSF phase would entail measurements of both phase coherence below the BKT temperature, and a finite jump discontinuity discriminating the pair from the independent superfluid phases.

\section*{Summary and Discussion } 
\label{sec:summary}

In summary, we have established the low-temperature phase diagram of bilayer IXs in the presence of finite density imbalance. We demonstrated that the predicted evolution of IX energy shifts as a function of density imbalance in various experimentally accessible parameter regimes allows distinguishing between single and paired superfluidity and probing the associated quantum critical phenomena separating the two phases. Via detailed analysis of realistic experimental parameters, we argue that our predictions can be validated experimentally in ongoing and near-future experiments of vertically stacked DQWs structures in GaAs- and TMD-based systems. 

We conclude our presentation by flagging several future lines of research, motivated by our results. First, a more accurate description of IX dynamics, beyond our simplified model, requires taking into account the effect of spatial confinement and disordering potentials. Sufficiently weak disorder is not expected to qualitatively modify our predictions, evaluated in the clean and uniform limit. Moreover, the strong dipolar interaction between IXs was predicted to effectively screen the in-plane disorder \cite{High_2009,Remeika_2009,Ivanov_2002}, suppressing the pinning effect of trapping potentials even at low IX densities. Interestingly, spatial confinement of IXs is expected to increase the BKT transition temperature.

A natural extension of the bilayer model is multilayer structures. In that regard, the additional layer degrees of freedom, beyond the bilayer limit, can give rise to complex symmetry breaking patterns such as zigzag states. Moreover, the expected effective mass of multilayer bound states will scale linearly with the number of layers. This, in turn, quenches the kinetic energy and enhances the effect of dipolar interactions. In the strong coupling regime, crystal phases are energetically favorable, enabling access to IX solids at relatively low IX densities. 

The above two modifications of the studied model can be treated via numerically exact QMC simulations, similar to the ones carried out in this work. We leave these intriguing lines of research to future studies.

\vspace*{1em}
\begin{acknowledgments}
We thank Ephraim Keren, Yotam Mazuz-Harpaz, Yoav Sagi, and Hadar Steinberg for instructive discussions. We also thank Eran Bernstein for contributing to the numerical code. M.Z. acknowledges support from the Dalia and Dan Maydan Fellowship and the Maria Proner-Pogonowska Scholarship for Women in Science. 
M.Z. and S.G. acknowledge support from the Israel Science Foundation Grant 1686/18 and from the US–Israel Binational Science Foundation (BSF) Grant 2018058. 
R.R. acknowledges support from the Israeli Science Foundation Grant 836/17 and from the NSF-BSF Grant 2019737.
\end{acknowledgments}

\bibliographystyle{apsrev4-2}
\bibliography{mybib_arxiv02}{}

\onecolumngrid

\appendix

\section{Additional QMC data}

\label{app:qmc}
\begin{figure*}[ht]
    \captionsetup[subfigure]{labelformat=empty}
	\subfloat[\label{fig:g11}]{}
    \subfloat[\label{fig:g12}]{}
    \subfloat[\label{fig:g12_r0_rd}]{}
    \includegraphics[width=1\textwidth]{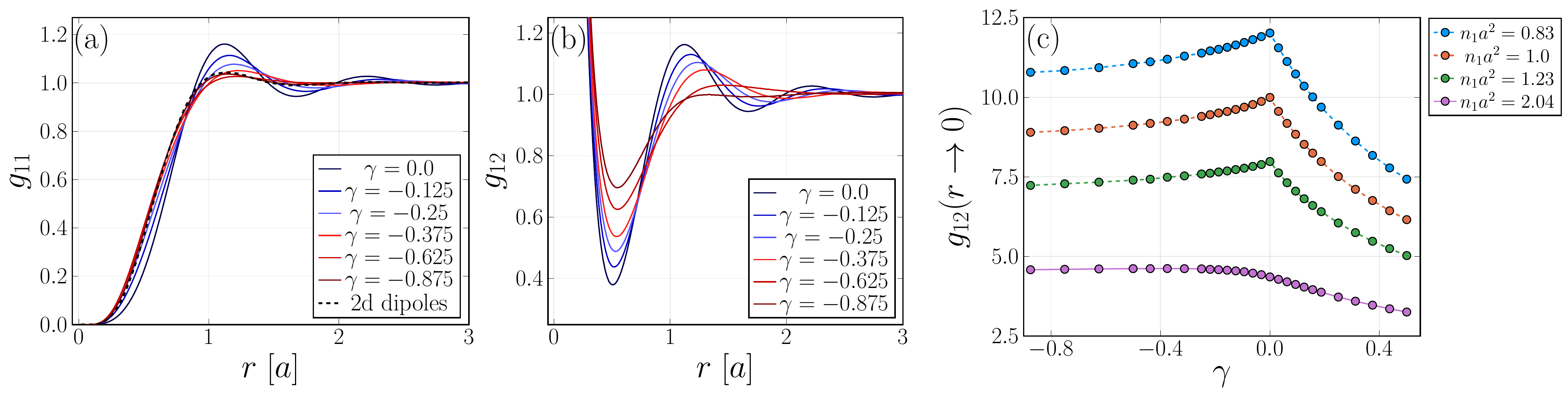}
    \caption{(a) Intralayer and (b) interlayer density-density correlation function plotted for several $\gamma$ values for $L_z=0.3[a]$ and $n_1=0.83[a^{-2}]$. The black dashed line in (a) corresponds to a single layer of two-dimensional dipolar bosons. (c) Zero separation limit of $g_{12}(r)$ as a function of $\gamma$ at $L_z=0.3[a]$. Density values $n_1$ residing in the PSF (2SF) phase at equal densities are shown by dashed (solid) lines. These results were obtained at $T=0.08 [E_0]$ and $N_1=32$.}
    \label{fig:gr}
\end{figure*}

In this section, we provide additional numerical results supporting the ones appearing in the main text.
We begin by computing the evolution of interlayer and intralayer density-density correlation functions for varying density imbalances. In \cref{fig:g11,fig:g12}, we plot $g_{11}(r)$ and $g_{12}(r)$, respectively, for several values of $\gamma$ at $L_z=0.3[a]$ and $n_1=0.83[a^{-2}]$, corresponding to the PSF phase at $\gamma=0$. We find that density imbalance smears the correlation peaks appearing both in $g_{11}(r)$ and $g_{12}(r)$. At large density imbalances, intralayer correlations eventually vanish and $g_{11}(r)$ approaches the monolayer result, shown with a dashed line in \cref{fig:g11}. To understand this effect, we note that paired dipoles are heavier and carry enlarged effective dipole moment, compared to independent dipoles. Hence, their kinetic energy is quenched and the repulsion between them is amplified, promoting spatial correlations as a precursor of a crystalline phase. In \cref{fig:g12_r0_rd}, we depict $g_{12}(r\to0)$ as a function of $\gamma$ for several density values of layer $\alpha=1$ at $L_z=0.3[a]$. Indeed, we find that for $n_1$ values within the PSF phase, dashed lines in \cref{fig:g12_r0_rd}, $g_{12}(r\to0)$ is maximal at $\gamma=0$ and decreases for greater imbalance. By contrast, for $n_1$ values in the 2SF phase, solid curve in \cref{fig:g12_r0_rd}, $g_{12}(r\to0)$ monotonically decreases with $\gamma$.

\begin{figure*}[ht]
    \captionsetup[subfigure]{labelformat=empty}
	\subfloat[\label{fig:Delta_scaling}]{}
    \subfloat[\label{fig:rho_scaling}]{}
    \includegraphics[width=0.8\textwidth]{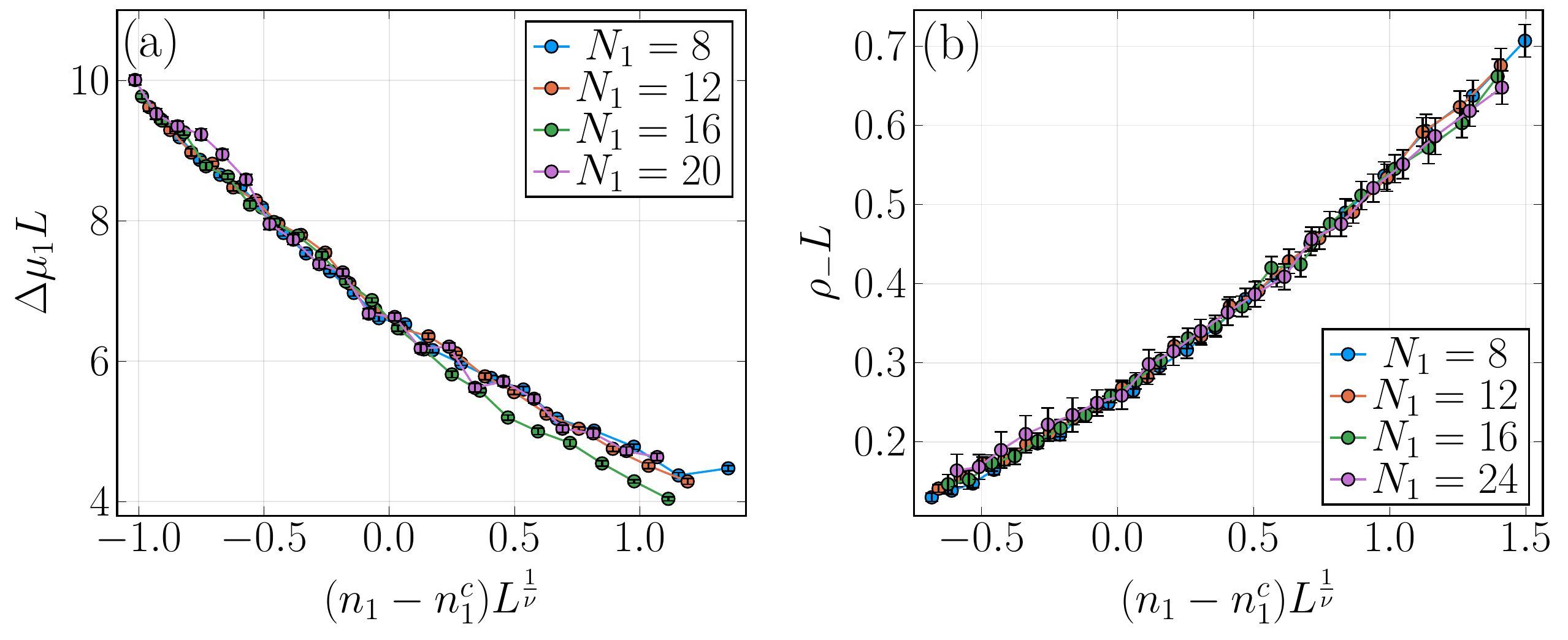}
    \caption{Universal curve collapse analysis of (a) the jump discontinuity, $\Delta\mu_1$, and (b) superfluid stiffness in the SCF channel, $\rho_-$, near the critical point $n_1^c$ for the PSF to 2SF transition. Different colors correspond to different system sizes. Results are shown for $T=0.08[E_0]$ and $L_z=0.4[a]$.}
    \label{fig:scaling}
\end{figure*} 

We now turn to study the universal scaling properties of $\Delta\mu_1$ and $\rho_-$ in the vicinity of the quantum critical point separating the PSF and 2SF phases at equal densities. Due to the layer exchange $\mathbb{Z}_2$ symmetry at density balance, the critical properties are expected to belong to the classical 3D XY universality class \cite{Safavi_Naini_2013}. 

To test the above prediction, we hypothesize the standard scaling form of the universal amplitudes $\Delta\mu_1 L ,\, \rho_- L \,\sim \tilde{f}_{\Delta/\rho}\qty(x)$, for some universal scaling functions $\tilde{f}_{\Delta/\rho}\qty(x)$. We assume relativistic dynamics and take the scaling variable $x=(n_1-n_1^c) L^{\frac{1}{\nu}}$. Here, $n_1^c$ is the critical density, $L$ is the linear system size, and $\nu$ is the correlation length exponent. In \cref{fig:Delta_scaling}, we plot $\Delta\mu_1 L$ as a function of the scaling variable, $x$, at $L_z=0.4[a]$. Indeed, we find an excellent curve collapse for $\nu_{\text 3D  XY}=0.671$ \cite{Burovski_2006} and $n_1^c=0.40(5)$ for curves corresponding to distinct system sizes. We obtain a similar result for $\rho_- L$, see \cref{fig:rho_scaling}, with $n_1^c=0.38(4)$. These results corroborates the predicted scaling behavior. 

\begin{figure*}[ht]
    \captionsetup[subfigure]{labelformat=empty}
	\subfloat[\label{fig:rho_p}]{}
    \subfloat[\label{fig:rho_m}]{}
   \includegraphics[width=0.8\textwidth]{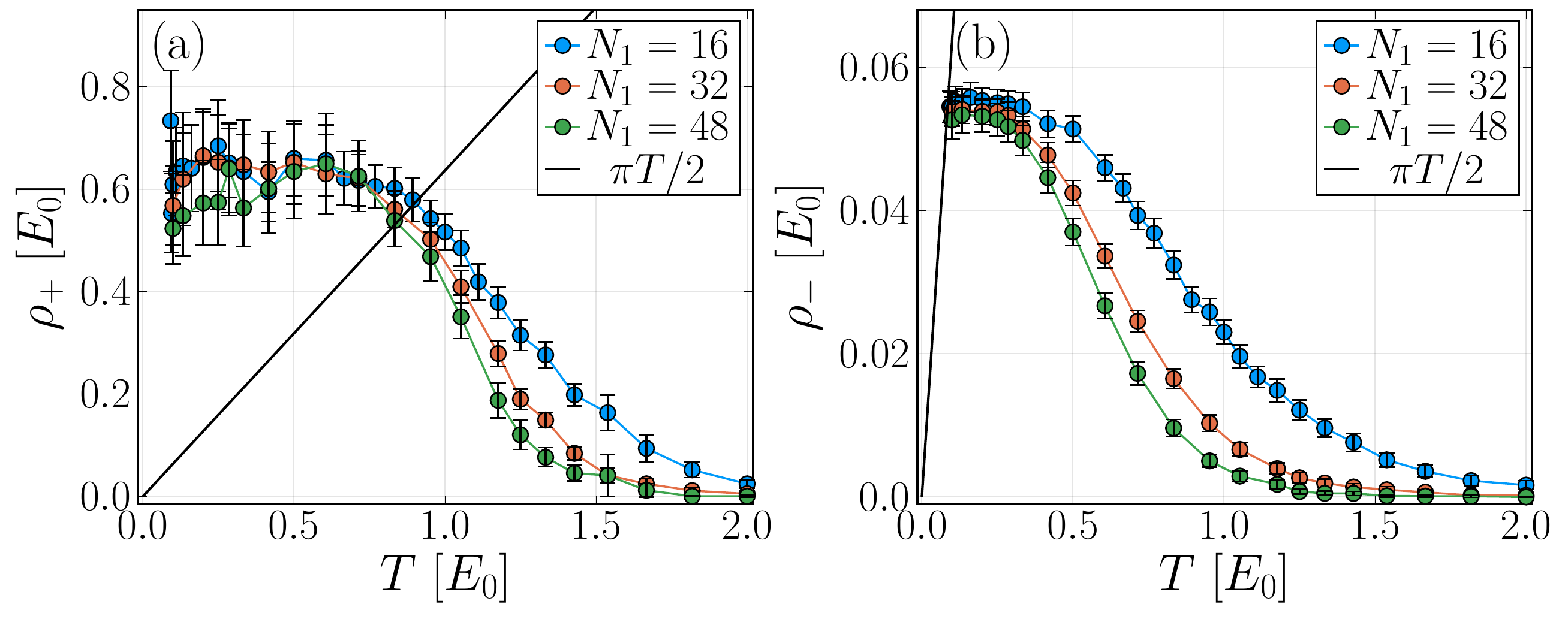}
    \caption{(a) $\rho_+$ and (b) $\rho_-$ as a function of the temperature, for several system sizes. The black line corresponds to the Nelson criterion $\frac{\pi}{2}T$. In the above figures we set $L_z=0.2[a]$, $n_1=1.23[a^{-2}]$ and $\gamma=0.125$.}
    \label{fig:rho_N_conv}
\end{figure*}

Next, in \cref{fig:rho_p,fig:rho_m}, we present a finite size scaling analysis of the two step BKT transitions, separating the normal to PSF and PSF to PSF+layer-($\alpha$) SF phases, at finite density imbalance $\gamma=0.125$, and for $L_z=0.2[a]$ and $n_1=1.23[a^{-2}]$. As in the main text, following the standard Nelson jump criterion \cite{Nelson_1977} $\rho_{\pm}(T=T_c)=\frac{\pi}{2}T_c$, we search for the  crossings of $\rho_{\pm}$ curves and the line $\frac{\pi}{2}T$. We find good agreement among crossing temperatures belonging to different systems sizes, allowing for an accurate determination of the BKT transition temperatures.

\begin{figure*}[ht]
    \captionsetup[subfigure]{labelformat=empty}
	\subfloat[\label{fig:E_conv_T}]{}
    \subfloat[\label{fig:rho_conv_T}]{}
    \includegraphics[width=0.80\textwidth]{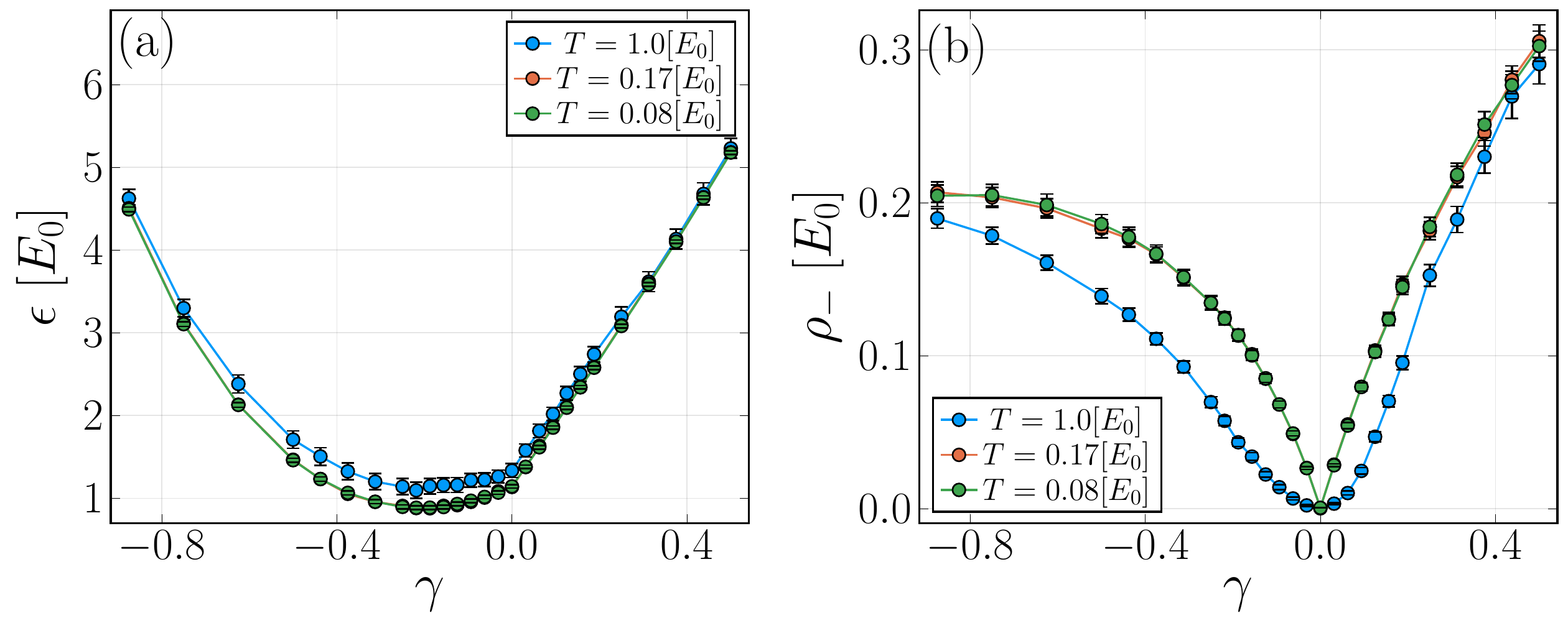}
    \caption{Zero temperature convergence of (a) $\epsilon$ and (b) $\rho_-$ as a function of $\gamma$. Here, we set $L_z=0.3[a]$, $n_1=0.83[a^{-2}]$ and $N_1=32$.}
    \label{fig:T_conv}
\end{figure*}

\begin{figure*}[ht]
    \captionsetup[subfigure]{labelformat=empty}
	\subfloat[\label{fig:E_conv_N}]{}
    \subfloat[\label{fig:rho_conv_N}]{}
    \includegraphics[width=0.80\textwidth]{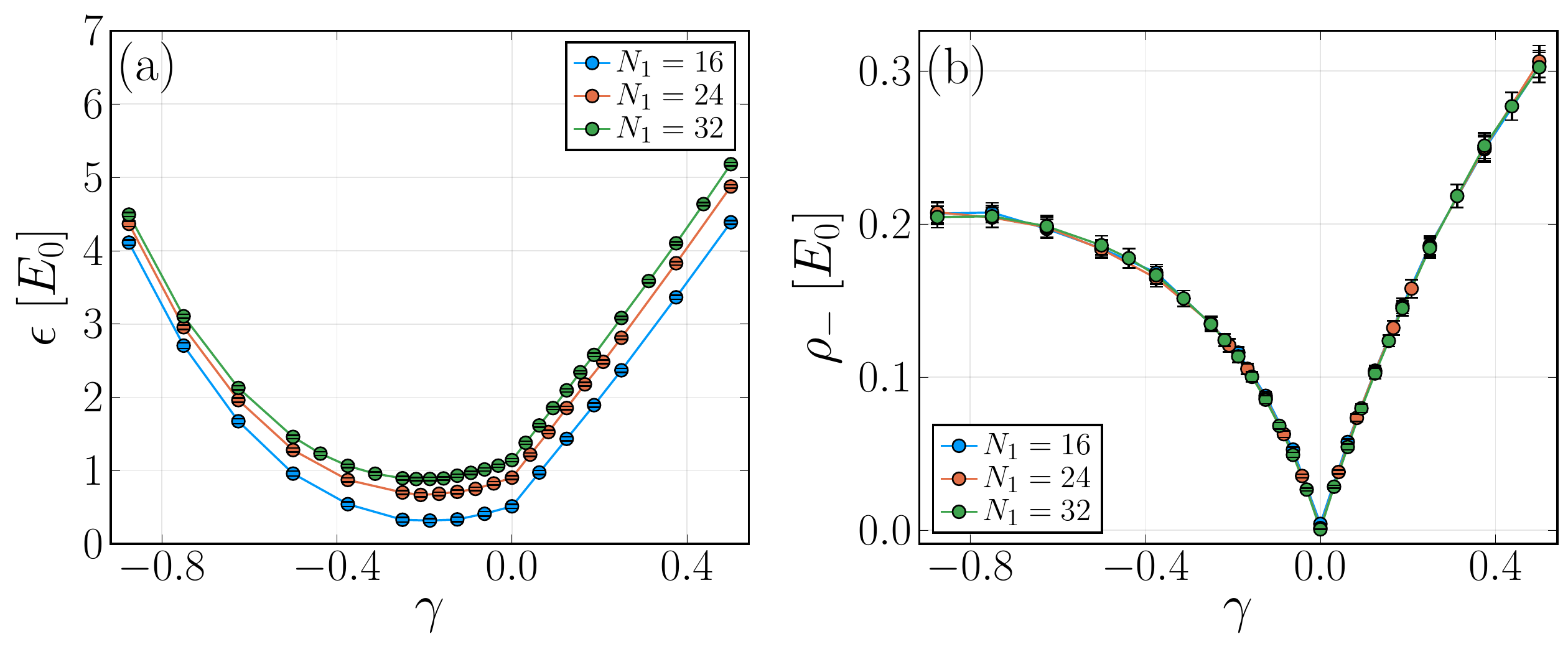}
    \caption{Finite size convergence of (a) $\epsilon$ and (b) $\rho_-$ as a function of $\gamma$ at $L_z=0.3[a]$, $n_1=0.83[a^{-2}]$ and $T=0.08[E_0]$.}
    \label{fig:N_conv}
\end{figure*}

We now proceed to monitor the finite size and finite temperature convergence of our numerical data. In \cref{fig:T_conv}, we depict the dependence of the total energy per particle, $\epsilon=E/(N_1+N_2)$, and superfluid stiffness, $\rho_-$, on density imbalance for $L_z=0.3[a]$, $n_1=0.83[a^{-2}]$, $N_1=32$ and a set of decreasing temperatures. We indeed observe convergence to the ground state value for the lowest temperature considered, $T=0.08[E_0]$. In \cref{fig:N_conv}, we again plot $\epsilon$ and $\rho_-$ as a function of $\gamma$ for an increasing range of system sizes at $L_z=0.3[a]$, $n_1=0.83[a^{-2}]$ and temperature $T=0.08[E_0]$. From the clear convergence of the curves, we can safely deduce that $N_1=32$ is sufficiently large to represent the thermodynamic limit result.

\begin{figure*}[htb]
    \includegraphics[width=0.45\textwidth]{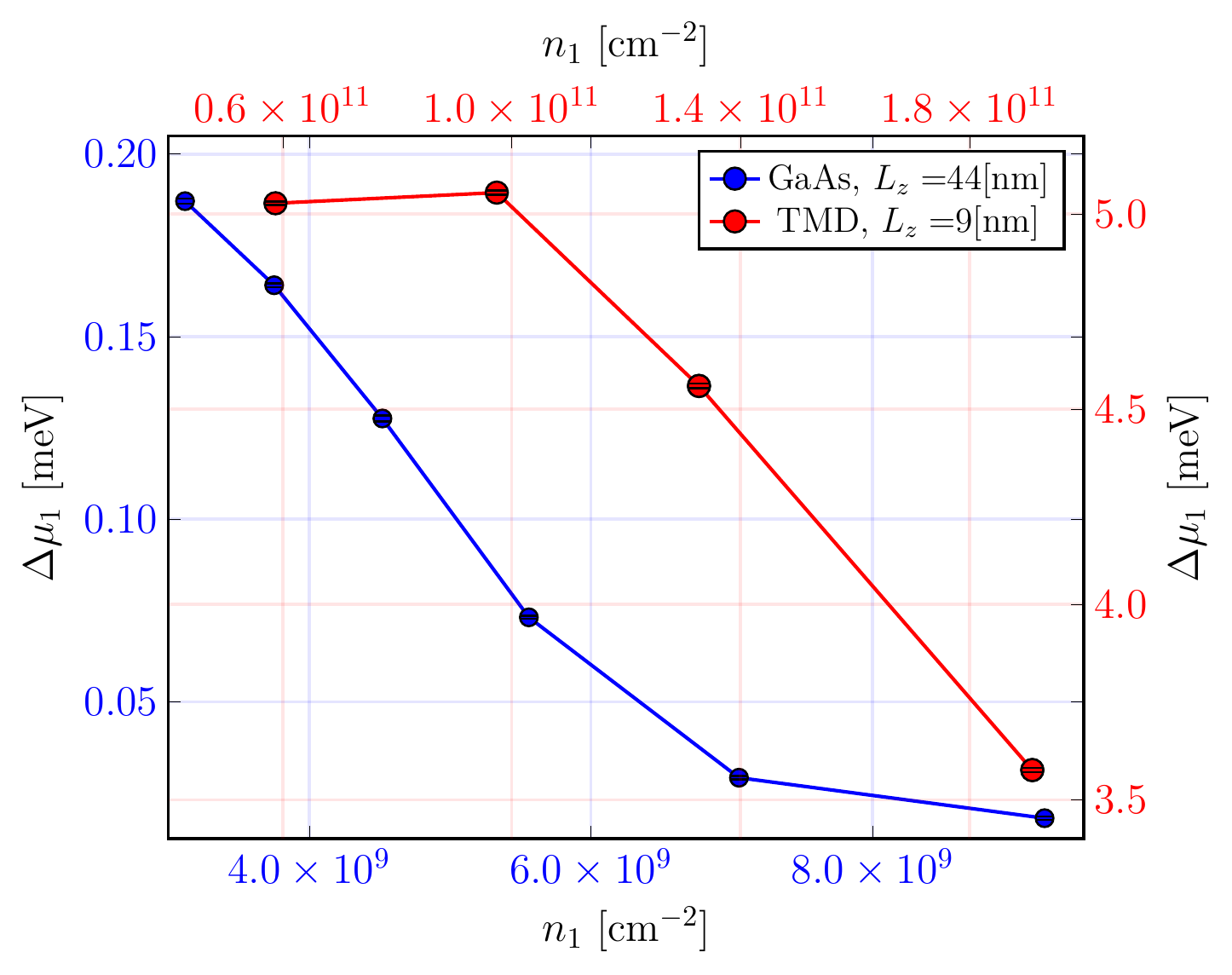}
    \caption{IX energy shifts as a function of IX densities. Blue and red colors correspond to the GaAs- and TMD- based settings described in the main text, respectively. The corresponding layer separations of both structures are denoted in the legend. These results we obtained in the ground state limit with $N_1=32$ and $N_2=16$ for the GaAs and TMD curves, respectively.}
    \label{fig:Fig_6_SM}
\end{figure*}

Lastly, we estimate the IX energy shift given by the jump discontinuity in the chemical potential, $\Delta\mu_1$, for several IX densities in the experimentally relevant regimes. We consider both GaAs and TMD based heterostructures with layer separations $L_z=44$ nm and $L_z=9$ nm, respectively, as proposed in the main text. We find that $\Delta\mu_1$ increases with decreasing density and saturates in the dilute limit, see \cref{fig:Fig_6_SM}. Moreover, the predicted energy shifts in TMD materials are considerably larger compared to the GaAs case, as discussed in the main text.

\section{Single IX solution}
\label{app:IX}

\begin{figure*}[t]
    \includegraphics[width=0.45\textwidth]{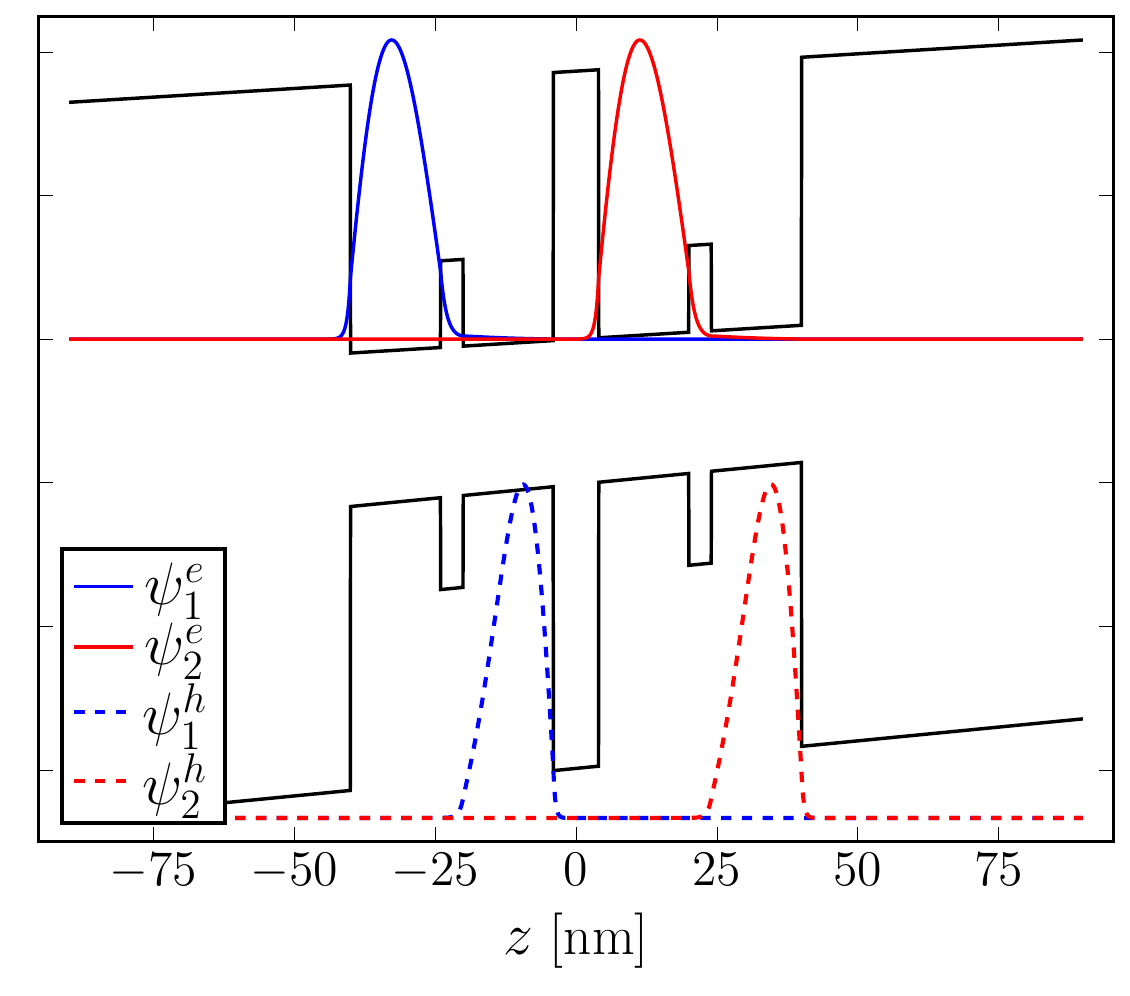}
    \caption{IX states in bilayer DQW potential in the presence of an external electric field $F_e=-2.5$ V/$\mu$m. The upper and lower black lines correspond to the conduction and valance band given by the bilayer DQW structure, respectively. The blue and red curves correspond to the electron (solid line) and hole (dashed line) wavefunctions at layer $\alpha=1,2$, respectively.}
    \label{fig:WF}
\end{figure*}

Here, we calculate the effective IX dipole sizes for the proposed GaAs-based bilayer DQW structure, see \cref{sec:exp_considerations} of the main text. To that end, we numerically solve the single particle Schr\"{o}dinger equation for the electron and hole, taking relative permittivity $\varepsilon=12.9$ and including an externally applied electric field $F_e$. In \cref{fig:WF}, we depict the electron and hole states, $\psi^{e(h)}_{\alpha}$, along the stacking direction, $z$, for electric field $F_e=-2.5$ V/$\mu$m. We find that the external field enhances IX dipole sizes in both layers relative to the zero voltage case, giving an increased effective dipole size $d=22$ nm.

\end{document}